\documentclass{article}

\usepackage{PRIMEarxiv}

\usepackage[utf8]{inputenc}
\usepackage[T1]{fontenc}
\usepackage{hyperref}
\usepackage{url}
\usepackage{booktabs}
\usepackage{amsfonts}
\usepackage{nicefrac}
\usepackage{microtype}
\usepackage{lipsum}
\usepackage{fancyhdr}
\usepackage{graphicx}
\graphicspath{{media/}}

\usepackage{comment}
\usepackage{booktabs}
\usepackage{multirow}
\usepackage{pgfplots}
\pgfplotsset{compat=1.18}
\usepackage{pgfplotstable}
\usepackage{xcolor}
\definecolor{brightmagenta}{HTML}{FF00FF}
\usepackage{colortbl}
\usepackage{siunitx}
\usepackage{caption}
\usepackage{float}
\usepackage{subfig}
\usepackage{makecell}
\usepackage{amssymb}
\usepackage{amsmath}

\fancypagestyle{firstpage}{%
  \fancyhf{}%
  \fancyfoot[C]{\footnotesize \textcopyright~2026 IEEE. Personal use is permitted but republication/redistribution requires IEEE permission. Accepted for publication in IEEE IECON.}%
}

\title{RDSE: Real-Time Directional Speech Enhancement Framework for Dual-Mic Commodity Smartphones}

\author{
Chinmay Deo \\
National Institute of Technology Jalandhar \\
Jalandhar, India \\
\texttt{chinmaydeo2004@gmail.com}
\And
Danish Mahajan \\
National Institute of Technology Jalandhar \\
Jalandhar, India \\
\texttt{mahajandanish581@gmail.com}
\And
Soudeep Ghoshal \\
Kalinga Institute of Industrial Technology \\
Bhubaneswar, India \\
\texttt{soudeepghoshal.sg@gmail.com}
\And
Himanshu Buckchash \\
IMC University of Applied Sciences \\
Krems, Austria \\
\texttt{himanshu.buckchash@imc.ac.at}
}

\begin{document}
\maketitle
\thispagestyle{firstpage}

\begin{abstract}
    Audio-visual zooming on mobile devices is fundamentally constrained by limited microphone count, compute resource, unknown array geometry, and weak assumptions on real-world acoustic conditions. We present RDSE: Real-Time Directional Speech Enhancement Framework, a hybrid pipeline that combines traditional signal processing and modern machine learning methods for dual-microphone low-end smartphones. Unlike tightly coupled AV systems that continuously use visual feeds or embeddings, RDSE decouples visual interaction from feature inference, using camera input solely to parameterize the steering vector for spatial filtering.
    The proposed system uses four modules:
    dereverberation network, a beamformer with a temporal convolutional network, and a MetricGAN based denoiser.
    Extensive evaluation on benchmark data and real-world recordings demonstrates effective zooming without precise geometry calibration. Comparative experiments show competitive performance to existing methods. Proposed system is rigorously tested for statistical validity, ablation study, latency profiling, and practicality through real-time live-streaming.

    \vspace{0.4em}
    \noindent \textbf{Project page \& live audio demos:} \textcolor{brightmagenta}{\url{https://danish-research.github.io/rdse-research.github.io/}}
\end{abstract}


\section{Introduction} \label{sec:introduction}

\textbf{Audio-visual zooming}~\cite{nair2019audiovisual} aims to isolate sound originating from a specific user-defined direction while suppressing competing interference and ambient noise, emulating the way human psychoacoustics leverage visual-auditory synchronization to address the \emph{cocktail party problem}~\cite{haykin2005cocktail}. On mobile devices, this capability is increasingly vital for applications such as high-fidelity conversational recording, remote conferencing, and assistive listening, reflecting the growing demand to capture clean audio from a focal point in acoustically crowded public environments.

Realising this on a commodity smartphone is fundamentally constrained by hardware. Dual-microphone arrays, standard on most modern devices, suffer from limited spatial resolution and pronounced directional ambiguity, rendering them highly sensitive to spatial covariance estimation errors and microphone geometry mismatch. Classical spatial filtering techniques, particularly \emph{Minimum Variance Distortionless Response (MVDR) beamforming}~\cite{capon1969mvdr}, provide a principled and interpretable framework for directional enhancement; however, their efficacy degrades substantially under these mobile constraints. End-to-end deep learning approaches~\cite{ephrat2018looking} have achieved state-of-the-art performance in source separation, yet they typically assume fixed array geometries, extended multi-microphone~\cite{nair2019audiovisual} configurations, or large matched training corpora, and frequently fail to generalise when deployed on unseen devices or acoustic environments that deviate from the training distribution. These limitations motivate a \textbf{hybrid spatial-neural design} that combines classical beamforming with modular neural refinement, while avoiding reliance on precise array calibration or continuous visual feature extraction.

In this paper, we present \textbf{RDSE} (\emph{Real-Time Directional Speech Enhancement}), a framework tailored specifically for dual-microphone smartphone arrays. RDSE is built around two core design principles: \textbf{one-time visual parameterization}, in which the camera contributes a single steering azimuth resolved from one user interaction rather than continuous feature extraction, and \textbf{geometrically-tolerant dual-microphone beamforming}, which operates without requiring calibrated array geometry. The proposed architecture cascades: (1) a U-Net based channel-wise dereverberation module \cite{ernst2018dereverb} to stabilise spatial covariance estimation under reverberant conditions; (2) an \emph{MVDR beamformer} \cite{capon1969mvdr} steered toward a user-defined direction; (3) a temporal convolutional network (\emph{Conv-TasNet})~\cite{luo2019convtasnet} for target-speaker extraction; and (4) a \emph{MetricGAN} based post-filter~\cite{fu2019metricgan} for residual noise suppression and perceptual enhancement.

\begin{figure}[t]
    \centering
    \includegraphics[width=0.7\linewidth]{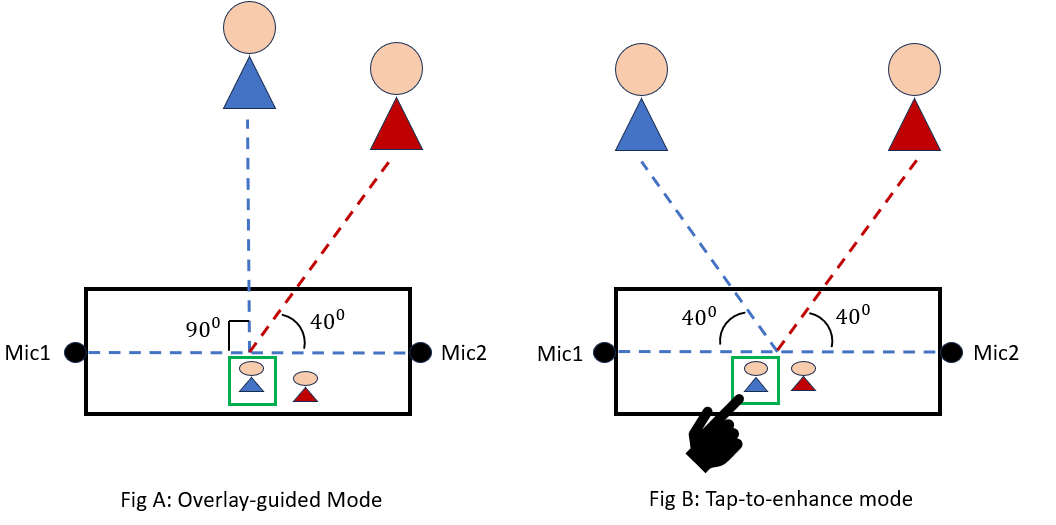}
    \caption{Two modes of RDSE}
    \label{fig:vista_zoom_modes}
\end{figure}

RDSE supports two interaction modes, illustrated in Figure~\ref{fig:vista_zoom_modes}. In \emph{overlay-guided mode}, the user aligns a reticle with the target, locking the beamformer to the broadside direction. In \emph{tap-to-enhance mode}, the user selects an arbitrary screen region, and the horizontal pixel coordinate is mapped to a steering azimuth relative to the device centre. Both modes share an identical audio backbone, ensuring that visual input is used exclusively for \textbf{intent specification} rather than feature-level inference. While RDSE supports both interaction paradigms, the present study focuses exclusively on the overlay-guided mode. The tap-to-enhance mechanism is retained as a conceptual extension of the framework.

The pipeline is evaluated on benchmark datasets and real-world recordings acquired via \emph{MATLAB Mobile}, demonstrating effective zooming without precise geometry calibration. A real-time live-streaming implementation further validates the system's practical viability on commercial mobile hardware, with computationally intensive neural stages offloaded to cloud based or offline processing. 

The main contributions are summarised as follows:

\begin{itemize}
    \item We propose \textbf{RDSE}, an audio zooming framework for constrained dual-microphone smartphone arrays that operates without visual embeddings or language dependence.

    \item We introduce a \textbf{one-time visual intent specification framework} in which the listening direction is defined via camera-guided alignment or tap based selection, explicitly decoupling visual control from audio inference for improved robustness under real-world conditions.

    \item We present \textbf{SpARZ100} (Spatial Audio for Realistic Zooming), a controlled evaluation dataset isolating anechoic and reverberant conditions for systematic benchmarking of mobile audio zooming systems.
\end{itemize}

\section{Related Works} \label{sec:related_works}

Audio-visual speech enhancement (AVSE) has demonstrated that visual cues significantly improve target speech extraction in multi-speaker settings. Early fusion based systems established the encoder-fusion-decoder paradigm, where facial or lip embeddings condition a spectrogram masking network~\cite{hou2018audio,ephrat2018looking}. Subsequent work improved visual representations using lip-reading-grade features~\cite{afouras2018conversation} and explored robustness under visual corruption via switching or variational formulations~\cite{sadeghi2020audio,sadeghi2021switching}. 

However, tightly coupled frame-level visual conditioning remains vulnerable to occlusion, pose variation, and domain shift, which are common in mobile scenarios~\cite{afouras2019my,sharma2023egocentric,kang2022impact}. To improve robustness, decoupled architectures such as Multimodal SpeakerBeam~\cite{ochiai2019multimodal} use fixed enrollment embeddings instead of continuous visual streams, resulting in more stable runtime behavior.

Multi-channel AV systems further combine spatial filtering with neural refinement~\cite{tan2020audio,xu2020neural}. While effective, these approaches assume calibrated arrays with four or more microphones and known geometry, limiting applicability to commodity smartphones. Camera-guided zooming has been explored in controlled multi-microphone prototypes~\cite{duong2017audio,nair2019audiovisual}, but these systems rely on calibrated arrays and continuous visual inference.
Mobile-efficient AV models have also been proposed~\cite{chuang2020lite,chuang2022improved,gogate2021towards,jung2024flowavse,li2025two}, focusing primarily on computational compression rather than geometric constraints. Even recent edge-oriented designs continue to depend on frame-level visual processing.

In contrast, \textbf{RDSE} addresses the dual-microphone low-to-mid range \textbf{smartphone} scenario. The system adopts one-time visual parameterization: a single user-guided azimuth estimate initializes a geometrically tolerant spatial stage, after which the audio pipeline operates independently of the visual stream.

\section{Signal Model}

We consider a dual-microphone recording scenario representative of commodity smartphone devices. Let $x_m(t)$ denote the continuous-time signal captured by the $m$-th microphone, $m \in \{1,2\}$. Each microphone signal is modeled as
\begin{equation}
x_m(t) = s(t) * h_{t,m}(t) + \sum_{k=1}^{K} i_k(t) * h_{i_k,m}(t) + n_m(t),
\end{equation}
where $s(t)$ denotes the target speech signal, $i_k(t)$ the $k$-th interferer, $n_m(t)$ additive noise, and $h_{t,m}(t)$ and $h_{i_k,m}(t)$ the corresponding channel impulse responses (CIRs).

After sampling, the discrete-time microphone signals are given by
\begin{equation}
x_m[n] =
\sum_{\ell=0}^{L_h-1} h_{t,m}[\ell]\, s[n-\ell]
+ \sum_{k=1}^{K} \sum_{\ell=0}^{L_h-1} h_{i_k,m}[\ell]\, i_k[n-\ell]
+ n_m[n],
\end{equation}
where $L_h$ denotes the effective channel length.

The CIR models both line-of-sight and multipath propagation and is expressed as
\begin{equation}
h_m(t) = \sum_{p=0}^{P} \alpha_p \, \delta(t - \tau_{p,m}),
\end{equation}
with $\alpha_p$ and $\tau_{p,m}$ denoting the attenuation and delay of the $p$-th propagation path to the $m$-th microphone.

A far-field propagation model is assumed, which is reasonable for typical conversational distances relative to the microphone spacing. Let $\mathbf{d} \in \mathbb{R}^2$ denote the inter-microphone displacement vector and $\theta$ the azimuth angle of arrival. The inter-microphone time difference of arrival (TDOA) is
\(\tau(\theta) = -\frac{\mathbf{d}^\top \mathbf{u}(\theta)}{c}\),

where $c$ is the speed of sound and $\mathbf{u}(\theta) = [\cos\theta \; \sin\theta]^\top$.

Accordingly, the frequency-domain steering vector for the two-microphone array is
\begin{equation}
\mathbf{a}(f,\theta) =
    \begin{bmatrix}
        1 \\
        e^{-j 2\pi f \tau(\theta)}
    \end{bmatrix}.
\end{equation}

This signal model underpins spatial covariance estimation and subsequent MVDR beamforming.

\begin{figure}[t]
    \centering
    \includegraphics[width=\linewidth]{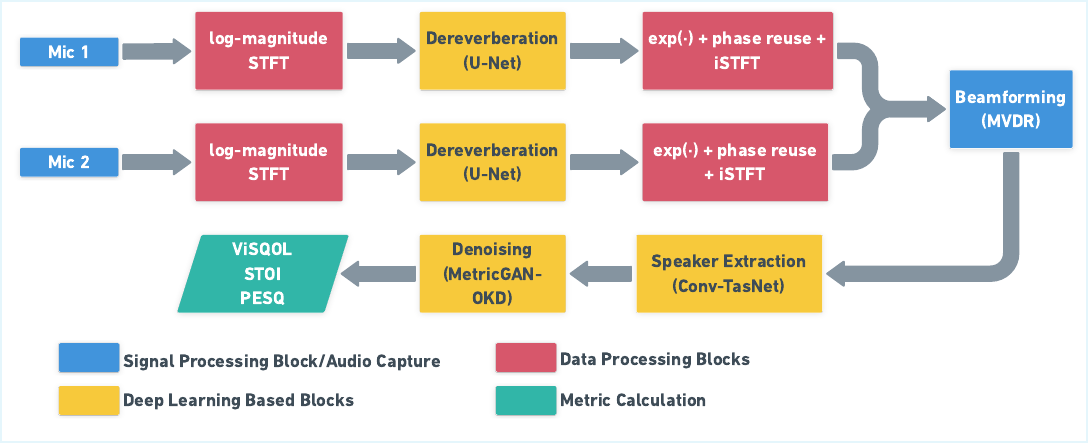}
    \caption{Block diagram of the proposed RDSE processing pipeline. Each microphone channel undergoes independent time-frequency transformation and U-Net based dereverberation prior to MVDR beamforming. The beamformer output is further refined using Conv-TasNet based target-speaker extraction and MetricGAN based speech enhancement.}
    \label{fig:vista_pipeline}
\end{figure}

\section{Methodology} \label{sec:methodology}

\subsection{RDSE Architecture}

The proposed RDSE pipeline integrates classical spatial filtering with modular, pretrained neural enhancement stages in a cascaded architecture. An overview of the complete system is shown in Fig.~\ref{fig:vista_pipeline}. The pipeline operates on dual-microphone audio captured by a smartphone device and produces a single-channel enhanced output corresponding to a user-specified listening direction.

\subsubsection{Dereverberation}

Dereverberation is applied independently to each microphone channel as a preprocessing step prior to beamforming to suppress late reflections and improve spatial covariance estimation for MVDR. We chose to keep dereverb before beamforming so that it can be turned off when not needed. It is applied channellwise to preserve inter-channel phase differences. In the deployable version of the pipeline, dereverberation is implemented as a user-controlled, toggleable module for heavy reverberant conditions (RT60 $> 0.7\,\text{s}$).

\begin{figure} [t]
    \centering
    \includegraphics[width=\linewidth]{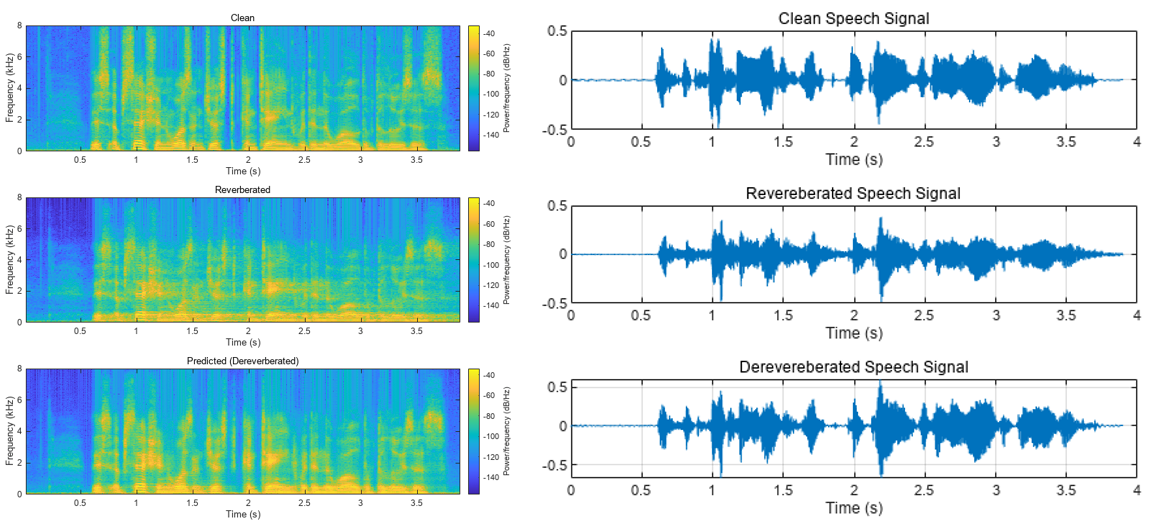}
    \caption{Effect of dereverberation on speech signals in the time-frequency and time domains \cite{ernst2018dereverb}.}
    \label{fig:dereverbnet_working}
\end{figure}

The dereverberation stage employs a pretrained U-Net based model (Fig.\ref{fig:dereverbnet_working}) operating in the time-frequency domain. The network takes the log-magnitude short-time Fourier transform (STFT) of the reverberant speech as input and predicts the corresponding dereverberated log-magnitude spectrum~\cite{ernst2018dereverb}.

Time-domain reconstruction is performed via inverse STFT reusing the phase of the reverberant input.

\noindent\textbf{STFT configuration:}
The short-time Fourier transform (STFT) is computed using a window length of $N=512$ samples with a peri    odic Hamming analysis window and $75\%  $ overlap ($384$ samples). An FFT length of $512$ is used, and a one-sided STFT representation is retained with $257$ frequency bins, where the highest-frequency bin is discarded, resulting in $256$ effective frequency bins.

\subsubsection{MVDR Beamforming}

Spatial filtering is performed using the minimum variance distortionless response (MVDR) beamformer~\cite{capon1969mvdr}. MVDR minimizes the output power of the beamformer while preserving a distortionless response in the desired look direction, thereby suppressing spatially incoherent interference and noise.

Let $\mathbf{X}(f,t) \in \mathbb{C}^{M}$ denote the $M$-channel STFT of the microphone signals at frequency bin $f$ and time frame $t$. The MVDR beamforming problem is formulated as
\begin{equation}
\begin{aligned}
\min_{\mathbf{w}(f)} \quad & \mathbf{w}^H(f)\mathbf{R}_{xx}(f)\mathbf{w}(f) \\
\text{s.t.} \quad & \mathbf{w}^H(f)\mathbf{a}(f,\theta_0) = 1 ,
\end{aligned}
\end{equation}
where $\mathbf{R}_{xx}(f)$ is the spatial covariance matrix of the multichannel observation, $\mathbf{a}(f,\theta_0)$ is the steering vector corresponding to the target direction of arrival (DoA) $\theta_0$, and $(\cdot)^H$ denotes the Hermitian transpose. For the overlay-guided mode, the target direction is fixed at $\theta_0 = 90^\circ$.

The closed-form solution to the MVDR optimization problem is given by
\begin{equation}
\mathbf{w}_{\text{MVDR}}(f) =
\frac{\mathbf{R}_{xx}^{-1}(f)\mathbf{a}(f,\theta_0)}
{\mathbf{a}^H(f,\theta_0)\mathbf{R}_{xx}^{-1}(f)\mathbf{a}(f,\theta_0)} .
\end{equation}

To improve numerical stability and robustness under finite-sample covariance estimation, diagonal loading \cite{xiao2017diagloading} is applied to the covariance matrix,

\begin{equation}
\mathbf{R}_{xx}(f) \leftarrow \mathbf{R}_{xx}(f) + \epsilon \, \mathrm{tr}\left(\mathbf{R}_{xx}(f)\right)\mathbf{I},
\end{equation}

where $\epsilon$ is a small positive constant and $\mathbf{I}$ denotes the identity matrix. This regularization mitigates ill-conditioning and improves beamformer performance in reverberant and noisy environments.

The beamformer output is computed as
\begin{equation}    
Y(f,t) = \mathbf{w}_{\text{MVDR}}^H(f)\mathbf{X}(f,t),
\end{equation}
followed by inverse STFT to obtain the time-domain beamformed signal.

\subsubsection{Speaker Extraction}

While MVDR beamforming provides spatial selectivity, residual interference and artifacts may persist due to geometry mismatch and nonstationary noise. To address this, the beamformer output is further refined using pretrained neural enhancement modules.
A temporal convolutional network (Conv-TasNet) \cite{luo2019convtasnet} is employed in a one-rest configuration to perform target-speaker extraction on the beamformed signal. The one-rest configuration focuses on extraction of dominant speaker.

\subsubsection{Denoising}

We use a pretrained MetricGAN \cite{fu2019metricgan} based speech enhancement network, which operates in the time-frequency domain and learns to improve perceptual quality metrics by estimating a magnitude mask while preserving phase. It suppresses non-speech components of audio.

\subsection{Real-Time Deployment and Validation}
\label{subsec:streaming}

Real-time RDSE uses a chunked streaming pipeline coupling MATLAB Mobile (capture) with desktop MATLAB (processing). Dual-channel audio is recorded at $16$~kHz via \texttt{mobiledev}. Audio is buffered and processed in $40$~ms blocks ($640$ samples) using a $512$-point Hann window with a hop size of $128$ samples ($75\%$ overlap), enabling low-latency overlap-add synthesis.

Spatial filtering is performed using a frequency-domain MVDR beamformer, with per-bin spatial covariance updated by exponential averaging,
\begin{equation}
\mathbf{R}_k^{(t)} = \alpha \mathbf{R}_k^{(t-1)} + (1-\alpha)\mathbf{x}_k^{(t)}\mathbf{x}_k^{(t)H},
\end{equation}
where $\alpha$ corresponds to a $0.5$~s time constant. Trace-proportional diagonal loading ensures numerical stability, and beamformer weights are recomputed per frame using a fixed, visually specified steering direction.

Beamformed spectra are reconstructed via inverse STFT and overlap--add synthesis to produce a continuous output. We refer to this as our ``streaming core". The computationally intensive neural modules are deferred to offline or cloud based post-processing.

\section{Experimental Setup}

\subsection{Dataset Description}

To evaluate the proposed pipeline under controlled acoustic conditions, we introduce \textbf{SpARZ100} (Spatial Audio for Realistic Zooming), a simulated dual-microphone dataset comprising two acoustic categories. Both categories simulate a cuboid room enclosure of dimensions $4.9\,\text{m} \times 4.9\,\text{m} \times 4.9\,\text{m}$, with a two-element omnidirectional microphone array positioned at the geometric center at a height of approximately $1.5\,\text{m}$. The interfering source is placed at an azimuth of $30^\circ$ relative to array broadside, and the signal-to-noise ratio (SNR) is fixed at $5\,\text{dB}$.

\begin{figure}[t]
    \centering
    \includegraphics[width=\linewidth]{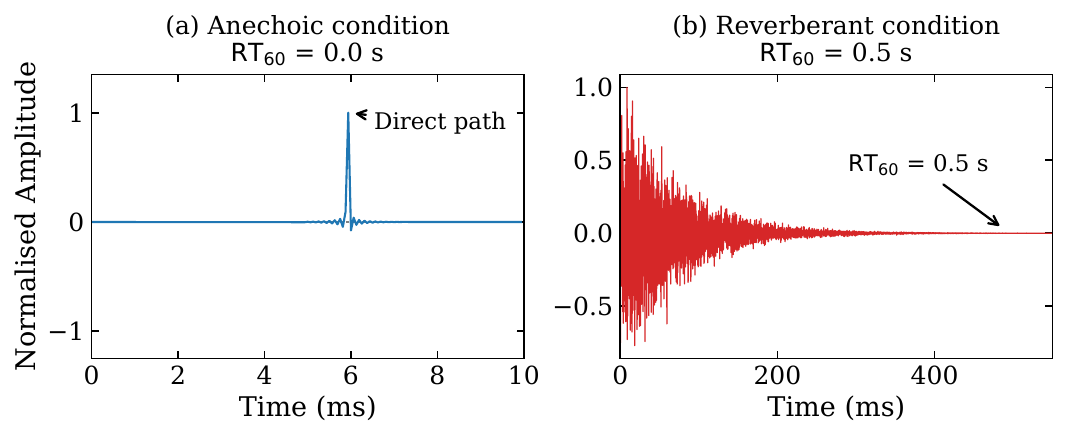}
    \caption{Comparison of room impulse responses (RIR) under two acoustic conditions. (a) Anechoic condition ($\text{RT60} = 0.0\,\text{s}$) showing only the direct path with no reflections. (b) Reverberant condition ($\text{RT60} = 0.5\,\text{s}$) showing the direct path followed by decaying reflections.}
    \label{fig:rir_comparison}
\end{figure}

The two categories differ exclusively in their acoustic condition. The first category represents an \textit{anechoic} condition, in which the reverberation time is set to $\text{RT60} = 0.0\,\text{s}$, meaning no reflections are present and the room impulse response reduces to a direct-path-only propagation model. The second category corresponds to a \textit{reverberant} condition, characterized by a reverberation time of $\text{RT60} = 0.5\,\text{s}$, with room reflections simulated via the image source method. RIRs corresponding to both RT60 values are visualised in Fig.\ref{fig:rir_comparison}. In both cases, the interfering source is placed at an azimuth of $30^\circ$ and the SNR is held constant at $5\,\text{dB}$. Each category comprises $50$ evaluation samples, yielding a total of $100$ samples across both conditions. This design enables a direct, controlled comparison of pipeline performance under anechoic versus reverberant acoustic environments, isolating reverberation as the sole varying factor. The target speech signals were drawn from the LibriSpeech corpus~\cite{7178964}, using male speaker utterances, while the interfering signals were sourced from the MUSAN dataset~\cite{snyder2015musanmusicspeechnoise}, using female speaker utterances.

\subsection{Dataset Characteristics}

\begin{figure}[t]
    \centering
    \includegraphics[width=\linewidth]{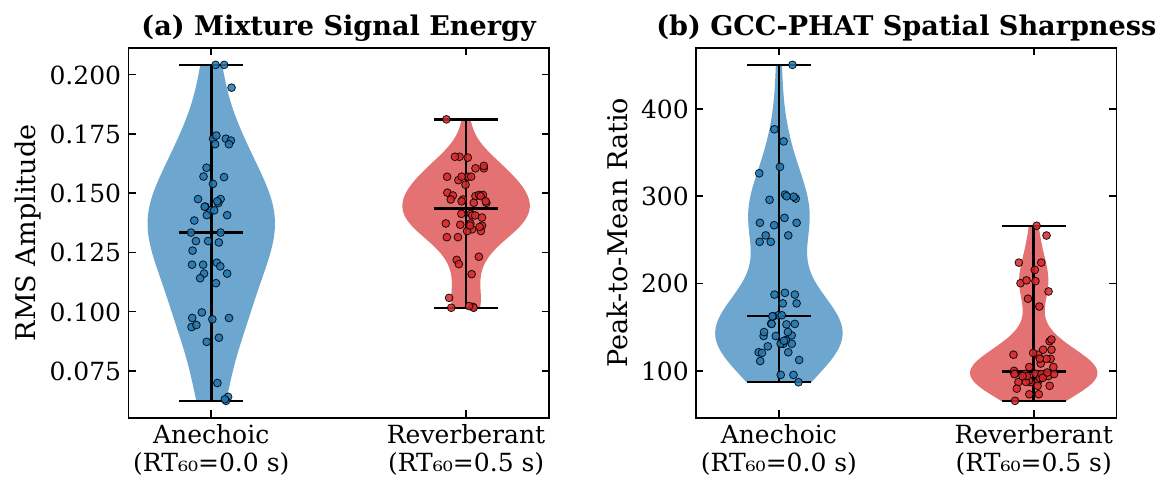}
    \caption{Distribution of (a) mixture signal energy (RMS amplitude) and (b) GCC-PHAT spatial sharpness (peak-to-mean ratio) across anechoic and reverberant conditions.}
    \label{fig:dist}
\end{figure}

\begin{figure}[t]
    \centering
    \includegraphics[width=\linewidth]{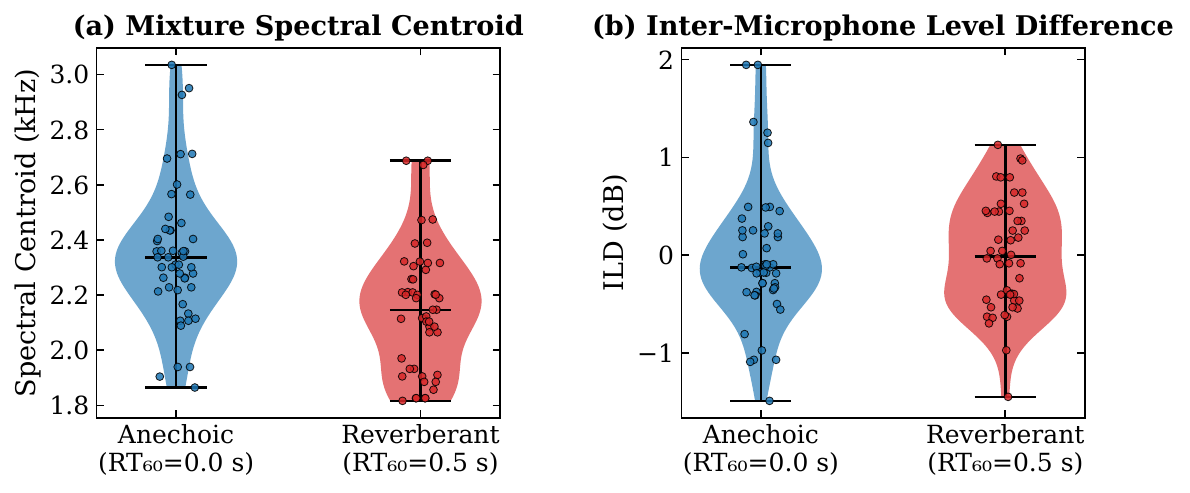}
    \caption{Distribution of (a) mixture spectral centroid and (b) inter-microphone level difference (ILD) across anechoic and reverberant conditions.}
    \label{fig:spectral}
\end{figure}

To provide insight into the acoustic properties of the evaluation dataset, four signal-level features were analysed across the anechoic ($\text{RT60} = 0.0\,\text{s}$) and reverberant ($\text{RT60} = 0.5\,\text{s}$) conditions. Figure~\ref{fig:dist} illustrates the distribution of mixture signal energy and GCC-PHAT spatial sharpness, while Figure~\ref{fig:spectral} shows the spectral centroid and inter-microphone level difference (ILD) across both conditions.

The anechoic samples exhibit higher variance in RMS energy, reflecting purely content-driven differences between audio files, whereas the reverberant samples cluster at a slightly higher energy level due to the additive contribution of room reflections. The GCC-PHAT peak-to-mean ratio, a measure of spatial localization confidence, is substantially higher and more spread in the anechoic condition, confirming that the direct-path-only propagation yields sharp, reliable TDOA estimates. Under reverberation, this ratio drops markedly, indicating that reflections smear the cross-correlation function and reduce localization certainty. The spectral centroid shifts downward in the reverberant condition, consistent with the frequency-dependent absorption of high-frequency energy by room reflections. Finally, the ILD remains tightly centred near $0\,\text{dB}$ in the anechoic condition but becomes more dispersed under reverberation, as multi-path arrivals introduce inconsistent level differences between the two microphones. Collectively, these distributions confirm that the two evaluation subsets are acoustically distinct and representative of the intended anechoic and reverberant scenarios.

\subsection{Real-Time Data Capture Environment}

Real-world validation experiments are conducted in a domestic indoor environment using two speakers positioned at azimuths of $90^\circ$ (male) and $40^\circ$ (female) relative to the array axis. Recordings are captured using a Samsung F24 smartphone. The exact microphone spacing of the device is unknown and is approximated as 12\,cm for beamforming and spatial processing analysis.

\section{Evaluation Metrics}

To comprehensively evaluate separation, intelligibility, and perceptual quality, we report three objective metrics: STOI, ViSQOL, and PESQ. These metrics collectively assess spatial separation performance, speech intelligibility, and perceptual audio quality.

\textbf{STOI.}
Speech intelligibility is measured using the Short-Time Objective Intelligibility (STOI) metric \cite{taal2010stoi}. STOI operates on short-time temporal envelopes derived from one-third octave band representations and produces a correlation based score in the range $[0,1]$, where higher values indicate better intelligibility.
STOI is computed between the clean reference and the final processed output.

\textbf{ViSQOL.}
Perceptual speech quality is further assessed using ViSQOL (Virtual Speech Quality Objective Listener) \cite{hines2012visqol}. ViSQOL is a spectro-temporal similarity based metric that compares time-frequency patches between reference and degraded signals. Scores lie in the range $[0,5]$, with higher values indicating improved perceptual quality.

\textbf{PESQ.}
We also report the Perceptual Evaluation of Speech Quality (PESQ) metric \cite{rix2001pesq}, which models human auditory perception to estimate speech quality. PESQ scores typically lie in the range $[-1, 4.5]$, with higher values corresponding to better perceptual quality.
Since PESQ is not natively available in our MATLAB evaluation environment, we compiled the reference implementation from source, generated a standalone executable, and developed a custom MATLAB wrapper to integrate it into our evaluation pipeline.

\section{Results and Discussion} \label{sec:results}

\subsection{Perceptual Quality and Intelligibility}

Table~\ref{tab:metrics} summarises the perceptual and intelligibility performance of the RDSE pipeline across the two evaluation conditions. The pipeline processes dual-microphone input through a cascaded architecture comprising MVDR beamforming, Conv-TasNet based speaker extraction, and MetricGAN speech enhancement. An optional U-Net dereverberation stage (DereverbNet) may be prepended to the pipeline prior to beamforming when operating in heavily reverberant environments.

\begin{table}[t]
  \caption{Perceptual quality and intelligibility metrics for RDSE}
  \label{tab:metrics}
  \centering
  \resizebox{.6\columnwidth}{!}{%
  \begin{tabular}{lccc}
    \toprule
    \textbf{Condition} & \textbf{STOI} & \textbf{ViSQOL} & \textbf{PESQ} \\
    \midrule
    Anechoic (RT$_{60}$ = 0.0\,s)    & 0.7754 & 2.0053 & 2.5755 \\
    Reverberant (RT$_{60}$ = 0.5\,s) & 0.5539 & 1.871  & 1.8185 \\
    \bottomrule
  \end{tabular}}
\end{table}

Under the anechoic condition (RT$_{60}$ = 0.0\,s), the pipeline achieves a STOI of 0.7754, a ViSQOL score of 2.0053, and a PESQ score of 2.5755. The processed output waveform closely tracks the structure of the clean reference, with the interfering speaker and background noise substantially suppressed. The relatively high PESQ score under this condition reflects the benefit of unambiguous spatial covariance estimation in the absence of reverberation, which allows the MVDR beamformer to steer accurately toward the target and effectively null the interferer located at 30$^\circ$.

Under the reverberant condition (RT$_{60}$ = 0.5\,s), performance degrades across all three metrics. This degradation is consistent with the known sensitivity of MVDR beamforming to spatial covariance estimation errors introduced by room reflections. Despite this, the processed waveform retains the temporal structure of the target speech, indicating that the downstream Conv-TasNet and MetricGAN stages provide partial compensation for the degraded beamformer output. The performance degradation under reverberation highlights the sensitivity of the overall pipeline to acoustic conditions. The effect of the U-Net dereverberation module is condition-dependent and is examined further in the ablation analysis in Section~\ref{subsec:ablation}.

\begin{figure}
\centering

\subfloat[Anechoic (RT$_{60}=0.0\,\mathrm{s}$)]{
    \includegraphics[width=\columnwidth]{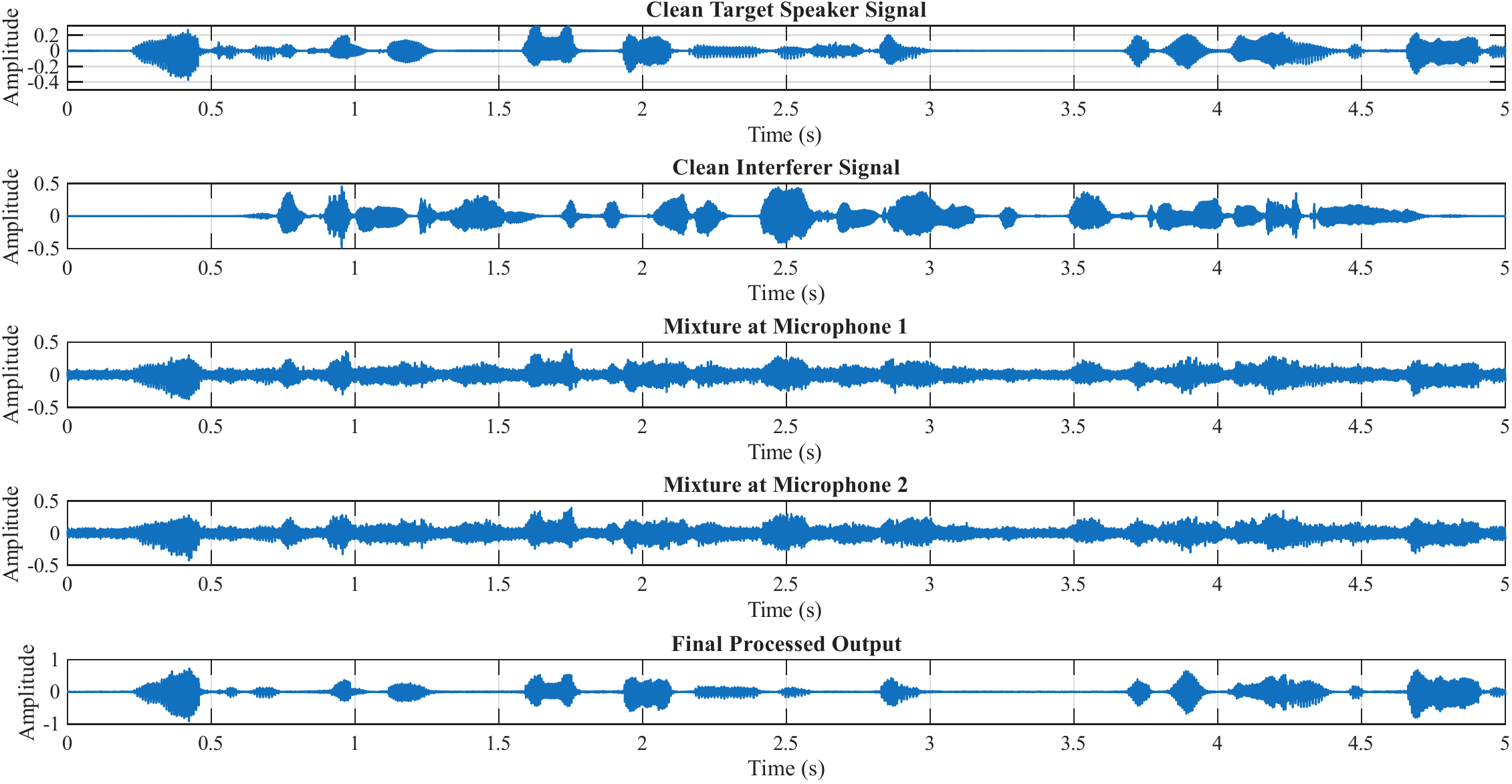}
    \label{fig:anechoic_wave}
}

\vspace{1em}

\subfloat[Reverberant (RT$_{60}=0.5\,\mathrm{s}$)]{
    \includegraphics[width=\columnwidth]{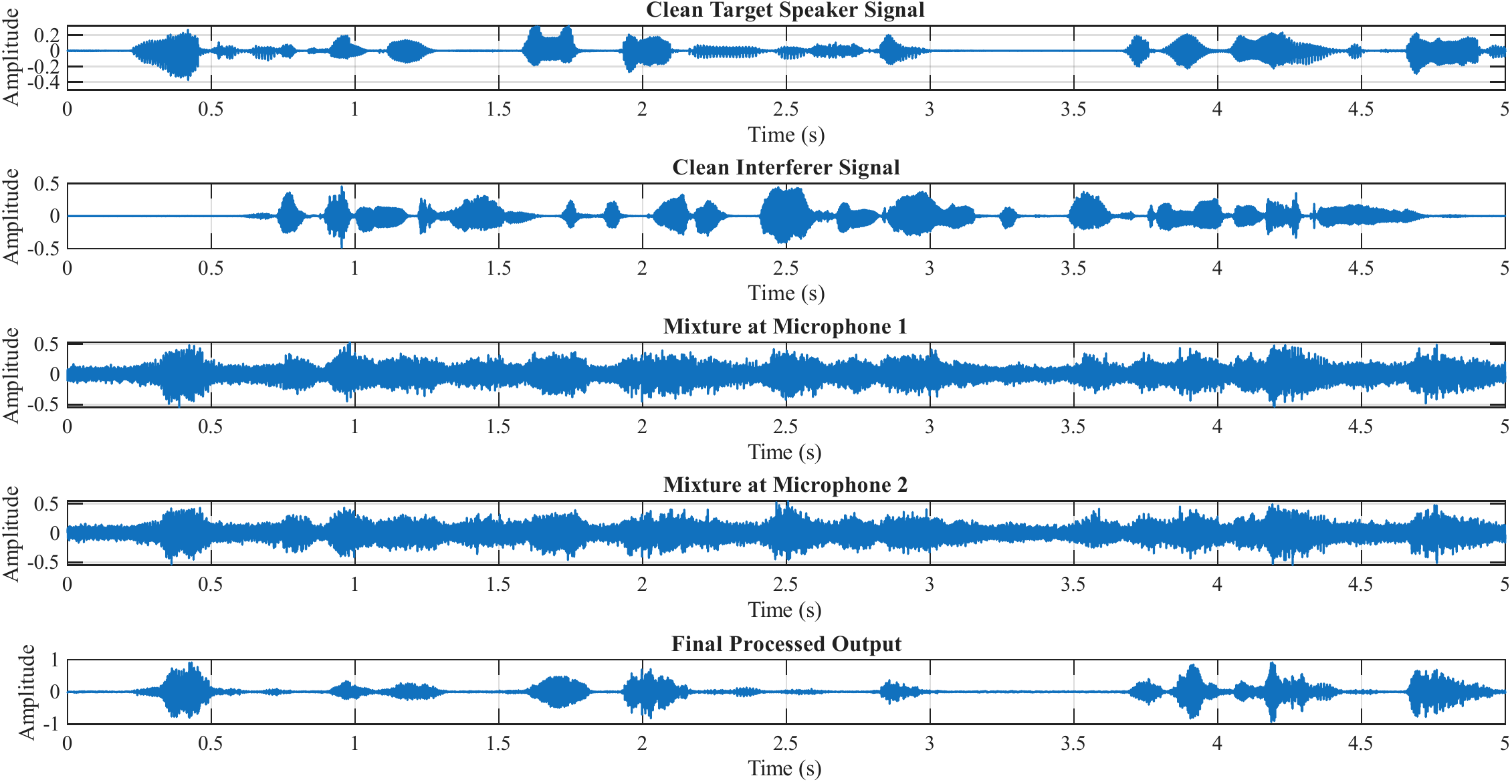}
    \label{fig:reverberant_wave}
}

\caption{Representative waveforms under anechoic and reverberant conditions. Each panel shows the clean target speech, the mixtures recorded at Mic\,1 and Mic\,2, and the final RDSE output.}
\label{fig:waveforms}
\end{figure}

Figures~\ref{fig:anechoic_wave} and~\ref{fig:reverberant_wave} show representative waveforms of the clean target, the mixture received at each microphone, and the final processed output under anechoic and reverberant conditions, respectively.

\subsection{Neural Backbone Selection}

To identify the most suitable neural speech extraction backbone for RDSE, we first evaluate three representative architectures: Conv-TasNet~\cite{luo2019convtasnet}, SepFormer~\cite{subakan2021attention}, and FasNet-TAC~\cite{luo2020end} under identical conditions on the proposed SpARZ100 benchmark. These models are evaluated independently as candidate speech extraction modules rather than complete directional speech enhancement systems. As shown in Table~\ref{tab:backbone}, SepFormer achieves the highest PESQ and ViSQOL scores, while Conv-TasNet exhibits mixed STOI performance across the two acoustic conditions. However, Conv-TasNet provides a more favorable balance between speech enhancement performance and computational efficiency, reducing inference time by approximately $3.8\times$ compared to SepFormer. Consequently, Conv-TasNet is selected as the neural speech extraction backbone for RDSE.

\begin{table}[t]
\caption{Neural backbone selection on SpARZ100}
\label{tab:backbone}
\centering
\setlength{\tabcolsep}{6pt}
\resizebox{0.9\columnwidth}{!}{%
\begin{tabular}{lccccc}
\toprule
\textbf{Method} & \textbf{STOI} & \textbf{ViSQOL} & \textbf{PESQ} & \textbf{Time (s)} & \makecell{\textbf{Real-time}\\\textbf{Feasible}} \\
\midrule
\multicolumn{6}{c}{\textit{Anechoic (RT$_{60}$ = 0.0\,s)}}\\
\midrule
Conv-TasNet~\cite{luo2019convtasnet} & \textbf{0.6169} & --- & 1.1999 & 6.05 & \checkmark\\
SepFormer~\cite{subakan2021attention} & 0.5015 & \textbf{1.5779} & \textbf{1.6239} & 22.79 & $\times$\\
FasNet-TAC~\cite{luo2020end} & 0.4494 & 1.5670 & 1.0780 & \textbf{3.50} & \checkmark\\
\midrule
\multicolumn{6}{c}{\textit{Reverberant (RT$_{60}$ = 0.5\,s)}}\\
\midrule
Conv-TasNet~\cite{luo2019convtasnet} & 0.4612 & --- & 1.0544 & 5.62 & \checkmark\\
SepFormer~\cite{subakan2021attention} & \textbf{0.4761} & \textbf{1.8253} & \textbf{1.2387} & 25.35 & $\times$\\
FasNet-TAC~\cite{luo2020end} & 0.4332 & 1.5856 & 1.0820 & \textbf{3.31} & \checkmark\\
\bottomrule
    \multicolumn{6}{l}{\footnotesize
    Inference time denotes the mean processing time for a 5\,s audio sample.}
\end{tabular}}
\end{table}

\subsection{Comparison with Prior Work}
RDSE is compared with prior audio-visual speech enhancement and directional speech extraction methods proposed by Nair \emph{et al.}~\cite{nair2019audiovisual}, Li \emph{et al.}~\cite{li2025two}, and Ephrat \emph{et al.}~\cite{ephrat2018looking}. Collectively, these methods address the common task of extracting a target speaker in the presence of competing speakers while representing three distinct paradigms: audio-visual guidance, multi-microphone beamforming, and lightweight dual-mic processing. Because these methods were evaluated on different datasets, microphone configurations and acoustic environments (e.g., AVSpeech, VoxCeleb2, and TCD-TIMIT), the results in Table~\ref{tab:comparison} are intended as qualitative rather than direct benchmark comparisons.

As shown in Table~\ref{tab:comparison}, RDSE achieves competitive speech intelligibility and perceptual quality while remaining feasible for real-time deployment on commodity dual-mic smartphones. Although Li \emph{et al.} report higher STOI values, their method leverages sparse overlapping speech and was evaluated under different experimental conditions. Likewise, the system of Nair \emph{et al.} employs a six-microphone array, whereas RDSE is designed for commodity dual-mic devices. Overall, RDSE provides a favorable trade-off between enhancement quality, computational efficiency, and deployment practicality.

\begin{table}[t]
\caption{comparison with prior work}
\label{tab:comparison}
\centering
\setlength{\tabcolsep}{5pt}
\resizebox{\columnwidth}{!}{%
\begin{tabular}{lccccccc}
\toprule
\textbf{Method} & \textbf{STOI} & \textbf{ViSQOL} & \textbf{PESQ} & \makecell{\textbf{Time}\\\textbf{(s)}} & \makecell{\textbf{Real-time}\\\textbf{Feasible}} & \makecell{\textbf{Mic}\\\textbf{Setup}} & \makecell{\textbf{Test}\\\textbf{Dataset}}\\
\midrule
\multicolumn{8}{c}{\textit{Anechoic (RT$_{60}$ = 0.0\,s)}}\\
\midrule
RDSE (Dereverb-off) & \textbf{0.7754} & 2.0053 & \textbf{2.5755} & 3.95 & \checkmark & 2-mic & SpARZ100\\
RDSE (Denoising-off) & 0.7356 & \textbf{2.0482} & 0.8789 & 10.11 & \checkmark & 2-mic & SpARZ100\\
Nair et al.~\cite{nair2019audiovisual} & 0.7100 & --- & 2.0400 & --- & $\times$ & 6-mic & AVSpeech\\
Li et al.~\cite{li2025two} & \textbf{0.8400} & --- & 2.0300 & --- & $\times$ & 2-mic & VoxCeleb2\\
Ephrat et al.~\cite{ephrat2018looking} & --- & --- & 2.4200 & --- & $\times$ & 1-mic & TCD-TIMIT\\
\midrule
\multicolumn{8}{c}{\textit{Reverberant (RT$_{60}$ = 0.5\,s)}}\\
\midrule
RDSE (Dereverb-off) & \textbf{0.5539} & 1.8710 & \textbf{1.8185} & 3.69 & \checkmark & 2-mic & SpARZ100\\
RDSE (Denoising-off) & 0.5391 & \textbf{1.9160} & 0.8527 & 6.21 & \checkmark & 2-mic & SpARZ100\\
Nair et al.~\cite{nair2019audiovisual} & 0.5900 & --- & 2.0000 & --- & $\times$ & 6-mic & AVSpeech\\
Li et al.~\cite{li2025two} & \textbf{0.6000} & --- & 1.5400 & --- & $\times$ & 2-mic & VoxCeleb2\\
\bottomrule
\multicolumn{8}{l}{\footnotesize Results from prior work are reported only for qualitative context due to differing datasets and evaluation protocols.}\\
    \multicolumn{8}{l}{\footnotesize
    Inference time denotes the mean processing time for a 5\,s audio sample.}
\end{tabular}}
\end{table}

\subsection{Ablation Studies} \label{subsec:ablation}

To isolate the contribution of each pipeline stage, we evaluate three ablated configurations against our proposed system (MVDR\,+\,Conv-TasNet\,+\,Denoising): the baseline comprising MVDR beamforming and Conv-TasNet only (MVDR\,+\,Conv-TasNet), the pipeline with the MetricGAN enhancement stage disabled (Denoising off), and the full pipeline with DereverbNet enabled. Results are reported in Table~\ref{tab:ablation}.

\begin{table}[t]
  \caption{Ablation study results under anechoic and reverberant conditions. Inference time is reported as mean seconds per sample.}
  \label{tab:ablation}
  \centering
  \resizebox{0.9\columnwidth}{!}{%
  \begin{tabular}{lcccc}
    \toprule
    \textbf{Configuration} & \textbf{STOI} & \textbf{ViSQOL} & \textbf{PESQ} & \textbf{Time (s)} \\
    \midrule
    \multicolumn{5}{c}{\textit{Anechoic (RT$_{60}$ = 0.0\,s)}} \\
    \midrule
    RDSE (Dereverberation-on) & 0.7356 & 2.0482 & 0.8789 & 10.11 \\
    RDSE (Dereverberation-off) & 0.7754 & 2.0053 & 2.5755 &  3.95 \\
    RDSE (Denoising-off) & 0.7416 & 2.1047 & 1.9645 & 14.85 \\
    RDSE (MVDR+Conv-TasNet) & 0.7890 & 2.0713 & 1.9106 &  1.38 \\
    \midrule
    \multicolumn{5}{c}{\textit{Reverberant (RT$_{60}$ = 0.5\,s)}} \\
    \midrule
    RDSE (Dereverberation-on) & 0.5391 & 1.9160 & 0.8527 &  6.21 \\
    RDSE (Dereverberation-off) & 0.5539 & 1.8710 & 1.8185 &  3.69 \\
    RDSE (Denoising-off) & 0.5592 & 1.9360 & 1.3434 & 14.73 \\
    RDSE (MVDR+Conv-TasNet) & 0.5681 & 1.9171 & 1.8676 &  1.56 \\
    \bottomrule
  \end{tabular}}
\end{table}

Under the anechoic condition, the baseline (MVDR\,+\,Conv-TasNet) achieves the highest STOI (0.7890) at the lowest inference cost (1.38\,s per sample), confirming that the spatial-neural core of the pipeline alone captures the majority of intelligibility gains. Adding MetricGAN (proposed) reduces STOI marginally to 0.7754 but delivers a substantial improvement in PESQ (2.58 vs.\ 1.91), reflecting the enhancer's ability to suppress residual non-speech components and improve perceptual output quality. ViSQOL remains broadly stable across all RDSE configurations (2.00--2.11), suggesting that overall spectro-temporal fidelity is preserved regardless of which stages are active.

Adding the denoising stage (Dereverb-off) substantially improves PESQ in the anechoic case while maintaining competitive STOI, yielding the best overall quality-intelligibility trade-off. Enabling dereverberation (Dereverb-on) provides modest gains in ViSQOL and competitive STOI, but its impact on PESQ is mixed across acoustic conditions and comes at a higher computational cost.

Under the reverberant condition, the proposed system (Dereverb off) achieves a PESQ of 1.82, competitive with the baseline (1.87), while the Denoising-off configuration attains the highest STOI (0.5592) at the cost of reduced PESQ (1.34), reflecting the inherent trade-off between intelligibility preservation and perceptual noise suppression introduced by the MetricGAN stage.

\subsection{Real-World Validation}

\begin{figure}[t]
  \centering
  \includegraphics[width=\linewidth]{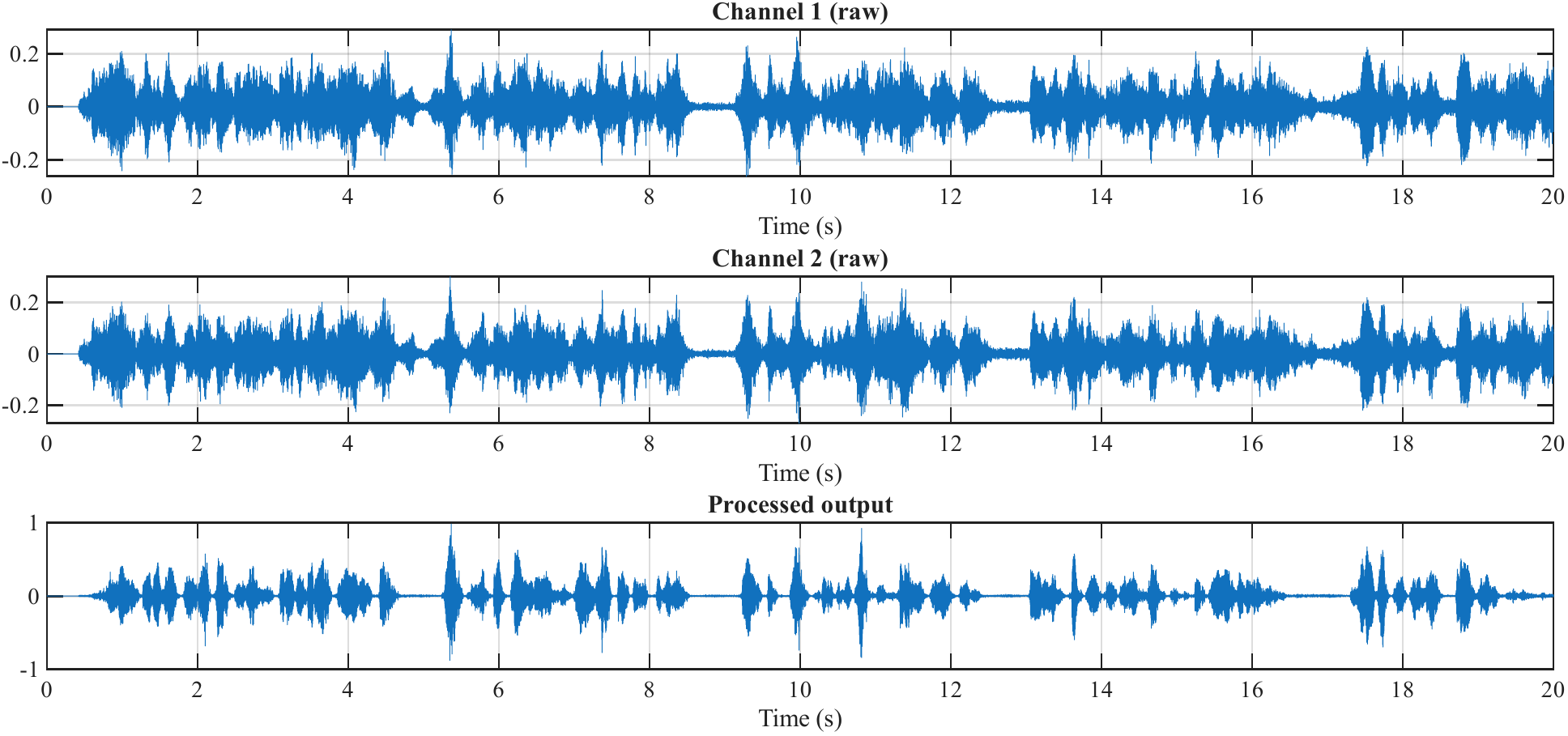}
  \caption{Time-domain waveforms from a real-world indoor
  recording (Samsung F24 smartphone). Top to bottom:  mixture at Mic\,1, mixture at Mic\,2,
  and final processed output.}
  \label{fig:real_world_waveforms}
\end{figure}

Figure~\ref{fig:real_world_waveforms} presents time-domain waveforms from a real-world indoor recording captured using a Samsung F24 smartphone, with a male target speaker at 90$^\circ$ and a female interferer at 40$^\circ$ relative to the array axis, and an approximated microphone spacing of 12\,cm. The raw microphone channels exhibit strong overlap between competing speakers along with environmental noise and room reflections. After processing, the output waveform demonstrates clear attenuation of the off-axis speaker while preserving the target's temporal structure. Despite unknown microphone geometry and uncontrolled acoustic conditions, the pipeline maintains stable separation, validating the practical robustness of the geometrically-tolerant beamforming design.

\subsection{Real-time Latency}
\label{subsec:latency}

Latency is end-to-end processing time and reported using the real-time factor (RTF), defined as the ratio of processing time to input duration. Table~\ref{tab:latency} summarizes the latency analysis.

\begin{table}[t]
\caption{End-to-end processing latency and real-time factor (RTF).}
\label{tab:latency}
\centering
\setlength{\tabcolsep}{4.5pt}
\renewcommand{\arraystretch}{1.05}
\resizebox{.5\columnwidth}{!}{%
\begin{tabular}{lccc}
\toprule
\textbf{Condition} & \textbf{Platform} & \makecell{\textbf{Input}\\\textbf{(s)}} & \makecell{\textbf{Latency}\\\textbf{(s)}} \\
\midrule
Live smartphone & MATLAB Mobile & 20.0 & 8.86 \\
\bottomrule
\end{tabular}}

\vspace{2pt}
\footnotesize{RTF: 0.351 (anechoic), 0.712 (reverberant). 
Streaming core operates in real time; cloud based neural inference = 8.86 s.}
\end{table}

On a live 20.09\,s recording processed on MATLAB Mobile, the streaming core requires 12.01\,s (RTF$_{\text{stream}}=0.60$). Per-frame (window-hop) statistics yield a median of 4.27\,ms and a 95\textsuperscript{th} percentile of 7.27\,ms, both below the hop duration $\Delta t=128/16000=8\,$ms, confirming deterministic real-time feasibility. The cloud based neural refinement (pre-trained, off-the-shelf MATLAB implementations) requires 8.86\,s (RTF$_{\text{cloud based}}=0.44$). This reflects unoptimized proof-of-concept performance without pruning, quantization, or distillation; standard model compression techniques are expected to further reduce latency. The combined runtime is 20.87\,s (RTF$_{\text{total}}\approx1.04$), demonstrating stable near-real-time operation on commodity hardware.

\section{Statistical Analysis}
\label{sec:statistical_analysis}

To rigorously compare the six methods under anechoic and reverberant conditions, we conducted a structured sequence of statistical tests: descriptive statistics to summarize performance, a global test to detect overall differences, post-hoc pairwise comparisons to identify specific differences, and a within-method test to quantify the impact of reverberation. All tests used a significance threshold of $\alpha = 0.05$.

To facilitate concise reporting across subsequent subsections and figures, the four RDSE pipeline configurations evaluated in this work are abbreviated as follows: \textbf{MC} = RDSE (MVDR\,+\,Conv-TasNet), the spatial-neural baseline comprising beamforming and Conv-TasNet only; \textbf{DR} = RDSE \textbf{(Dereverberation-off)}, the proposed system comprising MVDR, Conv-TasNet, and MetricGAN enhancement without dereverberation; \textbf{ES} = RDSE (Denoising-off), the proposed pipeline with MetricGAN disabled; and \textbf{FP} = RDSE (\textbf{Dereverberation-on}), the full pipeline with DereverbNet prepended.

\subsection{Descriptive Statistics and Confidence Intervals}
\label{subsec:summary_stats}

\begin{figure}
\centering
\includegraphics[width=0.7\linewidth]{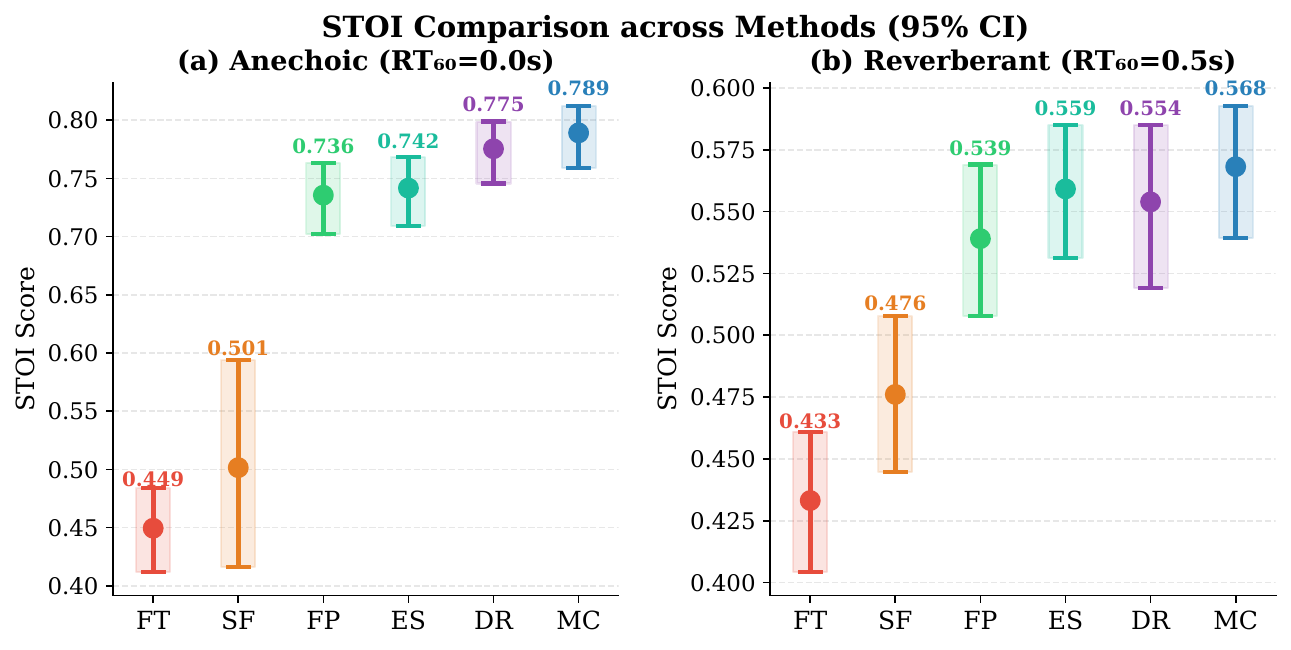}
\vspace{1em}
\includegraphics[width=0.7\linewidth]{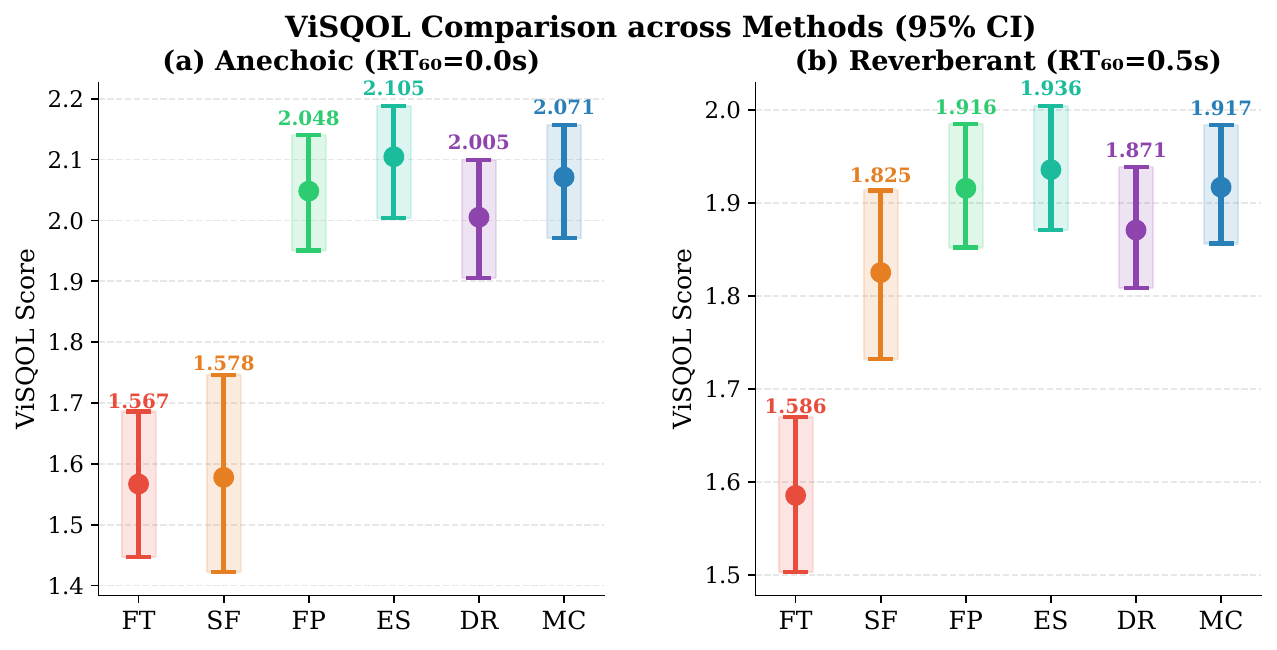}
\vspace{1em}
\includegraphics[width=0.7\linewidth]{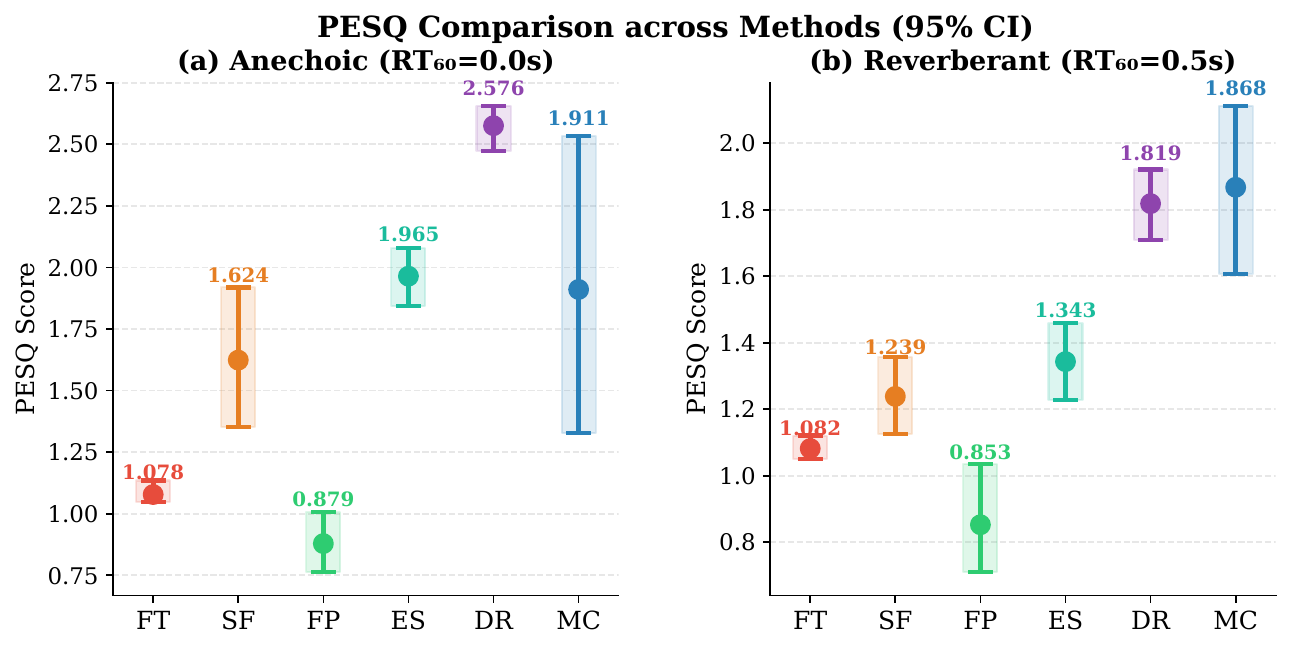}
\caption{Bootstrap 95\% confidence intervals for (a) STOI, (b) ViSQOL, and (c) PESQ. Anechoic ($\mathrm{RT}_{60}=0.0\,\mathrm{s}$) results are shown on the left and reverberant ($\mathrm{RT}_{60}=0.5\,\mathrm{s}$) on the right. Points denote mean scores and horizontal bars denote confidence bounds. Non-overlapping intervals indicate statistically significant separation. Abbreviations: FT = FasNetTAC, SF = SepFormer, FP = RDSE (Dereverberation-on), MC = RDSE (MVDR+Conv-TasNet), DR = RDSE (Dereverberation-off), ES = RDSE (Denoising-off). Exact values are given in the supplementary material.}
\label{fig:ci_plot_all}
\end{figure}

Figure~\ref{fig:ci_plot_all} visualises and reports the mean scores and 95\% bootstrap confidence intervals (CIs) for each method under both acoustic conditions, computed from 1000 bootstrap resamples. Bootstrap CIs are preferred here because they make no distributional assumptions, providing a robust estimate of uncertainty for non-normal data.

\textbf{RDSE (MVDR + Conv-TasNet)} achieves the highest STOI in both conditions (anechoic: $M = 0.789$, 95\% CI $[0.759, 0.812]$; reverberant: $M = 0.568$, 95\% CI $[0.539, 0.593]$). For perceptual quality, \textbf{RDSE (Dereverberation-off)} leads on PESQ in both conditions (anechoic: $M = 2.576$; reverberant: $M = 1.819$). \textbf{RDSE (Dereverberation-on)} yields unexpectedly low PESQ scores in both conditions ($M \approx 0.88$), suggesting that the full combination of processing stages may introduce perceptual artifacts. All RDSE variants exhibit a substantial STOI drop of approximately 0.20 points from anechoic to reverberant, while the baselines show minimal change due to their consistently low performance in both conditions.

\subsection{Across-Method Comparison Using Friedman Test}
\label{subsec:friedman}

To determine whether the six methods produce statistically different quality scores across the same set of audio samples, a Friedman test was applied separately for each metric-condition combination. The Friedman test is the non-parametric analogue of repeated-measures ANOVA, appropriate here because the same 50 samples are processed by all methods and the normality assumption has been shown to be violated. Table~\ref{tab:friedman} presents the test statistic $\chi^2(5)$, the $p$-value, and Kendall's $W$ as a measure of effect size.

\begin{table}[t]
\centering
\caption{Friedman test results for all metric-condition combinations. All tests are highly significant ($p \ll 0.001$), confirming that the six methods are not statistically equivalent. Kendall's $W$ reports the effect size; PESQ yields the highest concordance, indicating the most consistent ranking differences across samples.}
\label{tab:friedman}
\resizebox{0.8\columnwidth}{!}{%
\begin{tabular}{llccc}
\toprule
\textbf{Metric} & \textbf{Condition} & \textbf{$\chi^2(5)$} & \textbf{$p$-value} & \textbf{Kendall's $W$} \\
\midrule
\multirow{2}{*}{STOI}
  & Anechoic    & 85.97 & $4.72{\times}10^{-17}$ & 0.040 \\
  & Reverberant & 70.83 & $6.87{\times}10^{-14}$ & 0.033 \\
\midrule
\multirow{2}{*}{ViSQOL}
  & Anechoic    & 83.55 & $1.51{\times}10^{-16}$ & 0.039 \\
  & Reverberant & 45.38 & $1.21{\times}10^{-8}$  & 0.021 \\
\midrule
\multirow{2}{*}{PESQ}
  & Anechoic    & 41.31 & $8.12{\times}10^{-8}$  & 0.091 \\
  & Reverberant & 68.26 & $2.35{\times}10^{-13}$ & 0.118 \\
\bottomrule
\end{tabular}%
}
\end{table}

\begin{table*}
\centering
\caption{Nemenyi post-hoc $p$-values for \textbf{STOI} pairwise comparisons. Values below 0.05 indicate a statistically significant difference. \colorbox{green!25}{Green}: $p < 0.05$; \colorbox{gray!20}{grey}: not significant. FT = FasNetTAC, SF = Sepformer, DR = RDSE (Dereverberation-off), ES = RDSE (Denoising-off), FP = RDSE (Dereverberation-on), MC = RDSE (MVDR + Conv-TasNet).}
\label{tab:nemenyi_stoi}
\resizebox{\textwidth}{!}{%
\begin{tabular}{lcccccc}
\toprule
\multicolumn{7}{c}{\textbf{STOI — Anechoic}} \\
\midrule
 & \textbf{FT} & \textbf{SF} & \textbf{DR} & \textbf{ES} & \textbf{FP} & \textbf{MC} \\
\midrule
\textbf{FT} & {---} & \cellcolor{green!25}$0.0056$ & \cellcolor{green!25}$1.70{\times}10^{-10}$ & \cellcolor{green!25}$4.28{\times}10^{-6}$ & \cellcolor{green!25}$0.0011$ & \cellcolor{green!25}$2.22{\times}10^{-16}$ \\
\textbf{SF} & \cellcolor{green!25}$0.0056$ & {---} & \cellcolor{green!25}$0.014$ & \cellcolor{gray!20}$0.072$ & \cellcolor{gray!20}$0.108$ & \cellcolor{green!25}$9.87{\times}10^{-6}$ \\
\textbf{DR} & \cellcolor{green!25}$1.70{\times}10^{-10}$ & \cellcolor{green!25}$0.014$ & {---} & \cellcolor{gray!20}$0.642$ & \cellcolor{gray!20}$0.481$ & \cellcolor{gray!20}$0.871$ \\
\textbf{ES} & \cellcolor{green!25}$4.28{\times}10^{-6}$ & \cellcolor{gray!20}$0.072$ & \cellcolor{gray!20}$0.642$ & {---} & \cellcolor{gray!20}$0.992$ & \cellcolor{green!25}$0.0099$ \\
\textbf{FP} & \cellcolor{green!25}$0.0011$ & \cellcolor{gray!20}$0.108$ & \cellcolor{gray!20}$0.481$ & \cellcolor{gray!20}$0.992$ & {---} & \cellcolor{gray!20}$0.054$ \\
\textbf{MC} & \cellcolor{green!25}$2.22{\times}10^{-16}$ & \cellcolor{green!25}$9.87{\times}10^{-6}$ & \cellcolor{gray!20}$0.871$ & \cellcolor{green!25}$0.0099$ & \cellcolor{gray!20}$0.054$ & {---} \\
\bottomrule
\end{tabular}%
\quad\quad
\begin{tabular}{lcccccc}
\toprule
\multicolumn{7}{c}{\textbf{STOI — Reverberant}} \\
\midrule
 & \textbf{FT} & \textbf{SF} & \textbf{DR} & \textbf{ES} & \textbf{FP} & \textbf{MC} \\
\midrule
\textbf{FT} & {---} & \cellcolor{green!25}$0.033$ & \cellcolor{green!25}$1.00{\times}10^{-6}$ & \cellcolor{green!25}$1.19{\times}10^{-11}$ & \cellcolor{green!25}$2.16{\times}10^{-4}$ & \cellcolor{green!25}$2.46{\times}10^{-10}$ \\
\textbf{SF} & \cellcolor{green!25}$0.033$ & {---} & \cellcolor{gray!20}$0.831$ & \cellcolor{green!25}$4.38{\times}10^{-4}$ & \cellcolor{gray!20}$0.412$ & \cellcolor{green!25}$0.0025$ \\
\textbf{DR} & \cellcolor{green!25}$1.00{\times}10^{-6}$ & \cellcolor{gray!20}$0.831$ & {---} & \cellcolor{gray!20}$0.981$ & \cellcolor{gray!20}$0.743$ & \cellcolor{gray!20}$0.915$ \\
\textbf{ES} & \cellcolor{green!25}$1.19{\times}10^{-11}$ & \cellcolor{green!25}$4.38{\times}10^{-4}$ & \cellcolor{gray!20}$0.981$ & {---} & \cellcolor{gray!20}$0.634$ & \cellcolor{gray!20}$0.872$ \\
\textbf{FP} & \cellcolor{green!25}$2.16{\times}10^{-4}$ & \cellcolor{gray!20}$0.412$ & \cellcolor{gray!20}$0.743$ & \cellcolor{gray!20}$0.634$ & {---} & \cellcolor{gray!20}$0.567$ \\
\textbf{MC} & \cellcolor{green!25}$2.46{\times}10^{-10}$ & \cellcolor{green!25}$0.0025$ & \cellcolor{gray!20}$0.915$ & \cellcolor{gray!20}$0.872$ & \cellcolor{gray!20}$0.567$ & {---} \\
\bottomrule
\end{tabular}%
}
\end{table*}

\begin{table*}
\centering
\caption{Summary of significant Nemenyi pairwise comparisons for ViSQOL and PESQ ($p < 0.05$). Dashes (---) indicate non-significant differences for that specific pair and condition. Abbreviations: FT = FasNetTAC, SF = Sepformer, DR = RDSE (Dereverberation-off), ES = RDSE (Denoising-off), FP = RDSE (Dereverberation-on), MC = RDSE (MVDR + Conv-TasNet).}
\label{tab:nemenyi_visqol_pesq_wide}
\resizebox{\textwidth}{!}{%
\begin{tabular}{@{} l *{7}{cc} @{}}
\toprule
 & \multicolumn{2}{c}{\textbf{FT vs DR}} & \multicolumn{2}{c}{\textbf{FT vs ES}} & \multicolumn{2}{c}{\textbf{FT vs FP}} & \multicolumn{2}{c}{\textbf{FT vs MC}} & \multicolumn{2}{c}{\textbf{SF vs DR}} & \multicolumn{2}{c}{\textbf{FP vs DR}} & \multicolumn{2}{c}{\textbf{FP vs MC}} \\
\cmidrule(lr){2-3} \cmidrule(lr){4-5} \cmidrule(lr){6-7} \cmidrule(lr){8-9} \cmidrule(lr){10-11} \cmidrule(lr){12-13} \cmidrule(lr){14-15}
\textbf{Metric} & Anec. & Rev. & Anec. & Rev. & Anec. & Rev. & Anec. & Rev. & Anec. & Rev. & Anec. & Rev. & Anec. & Rev. \\
\midrule
\textbf{ViSQOL} & $<0.001$ & ---      & $<0.001$ & $0.012$ & $<0.001$ & $0.031$ & $<0.001$ & ---     & ---     & --- & ---      & ---      & --- & --- \\
\vspace{4pt}
\textbf{PESQ}   & $<0.001$ & $<0.001$ & $0.003$  & ---     & ---      & ---     & ---      & $0.004$ & $0.018$ & --- & $<0.001$ & $<0.001$ & --- & $0.023$ \\
\bottomrule
\end{tabular}%
}
\end{table*}

All six Friedman tests are highly significant ($p \ll 0.001$), conclusively rejecting the null hypothesis of method equivalence across every metric and condition. PESQ yields the highest concordance values ($W = 0.091$ anechoic, $W = 0.118$ reverberant), indicating that perceptual quality differences between methods are the most consistent across samples. Since the Friedman test is significant for all combinations, post-hoc pairwise comparisons are warranted.

\subsection{Pairwise Post-Hoc Comparisons Using Nemenyi Test}
\label{subsec:nemenyi}

The Nemenyi test was applied to identify which specific method pairs differ significantly, controlling the family-wise error rate across all 15 possible pairs simultaneously. Table~\ref{tab:nemenyi_stoi} presents the $p$-values for STOI pairwise comparisons under each condition, and Table~\ref{tab:nemenyi_visqol_pesq_wide} summarises significant pairs for ViSQOL and PESQ.

Under \textbf{anechoic conditions}, all four RDSE variants significantly outperform both FasNetTAC and Sepformer on STOI (all $p < 0.015$), with RDSE (MVDR + Conv-TasNet) achieving the strongest separation from both baselines. Among RDSE variants, RDSE (MVDR + Conv-TasNet) is significantly superior to RDSE (Denoising-off) ($p = 0.0099$), while no other intra-RDSE differences reach significance. Under \textbf{reverberant conditions}, RDSE variants again significantly outperform both baselines, but all intra-RDSE pairwise comparisons become non-significant, indicating that reverberation narrows the performance gap among pipeline configurations.

\subsection{Impact of Reverberation: Wilcoxon Signed-Rank Test}
\label{subsec:wilcoxon}

The Wilcoxon signed-rank test was applied as a paired comparison between anechoic and reverberant scores for the same 50 samples processed by each method. The null hypothesis states that the median performance under anechoic and reverberant conditions is equal. The rank-biserial correlation $r$ is reported as an effect size, where $|r| < 0.3$, $0.3 \leq |r| < 0.5$, and $|r| \geq 0.5$ correspond to small, medium, and large effects respectively. Positive $r$ indicates anechoic performance is higher; negative $r$ indicates reverberant performance is higher.

\textbf{FasNetTAC} shows no significant degradation under reverberation across any metric ($p > 0.05$, $|r| < 0.1$), reflecting a floor effect rather than robustness. Similarly, \textbf{Sepformer} shows no significant STOI change, and its ViSQOL score \emph{improves} significantly under reverberation ($p = 0.010$, $r = -0.429$), possibly because reverberant energy masks processing artifacts in its anechoic output. In contrast, \textbf{all four RDSE variants} exhibit statistically significant and very large STOI degradations under reverberation (all $p < 10^{-9}$, $r > 0.87$), a consequence of their strong anechoic performance creating a large gap that reverberation partially collapses. The largest PESQ degradation is observed in \textbf{RDSE (Dereverberation-off)} ($p < 10^{-11}$, $r = 0.944$), highlighting the condition-dependent impact of reverberation across the different RDSE configurations. \textbf{RDSE (Dereverberation-on)} is the only RDSE variant showing no significant PESQ change ($p = 0.632$), again due to a floor effect ($M \approx 0.88$ in both conditions).

\section{Conclusion} \label{sec:conclusion}

We introduced \textbf{RDSE}, a geometry-tolerant audio zooming framework for constrained dual-microphone smartphones that combines one-time visual intent specification with hybrid spatial-neural processing. The system operates without precise array calibration, continuous visual conditioning, or speaker/language dependence, making it practical for commodity devices.
Evaluations on the proposed \textbf{SpARZ100} dataset and real smartphone recordings demonstrate consistent intelligibility gains and competitive perceptual quality relative to strong neural baselines, while maintaining real-time feasibility. Ablation and non-parametric statistical analyses further validate the contribution of each module and the significance of observed improvements. Together, these results establish RDSE as a deployable and systematically evaluated solution for mobile audio zooming.

\bibliographystyle{unsrt}  
\bibliography{references}

\end{document}